\documentclass[aps,prl,reprint,superscriptaddress,twocolumn,twoside]{revtex4-1}

\usepackage{graphicx}
\usepackage{amsmath,amssymb}

\begin{document}

\title{Sub-kHz linewidth narrowing of a mid-infrared OPO idler frequency \\ by direct cavity stabilization}

\author{I.~Ricciardi}
\email{Corresponding author: iolanda.ricciardi@ino.it}
\author{S.~Mosca}
\author{M.~Parisi}
\author{P.~Maddaloni}
\author{L.~Santamaria}
\affiliation{CNR-INO, Istituto Nazionale di Ottica,
Via Campi Flegrei 34, 80078 Pozzuoli (NA), Italy}
\author{P.~De~Natale}
\affiliation{CNR-INO, Istituto Nazionale di Ottica,
largo E. Fermi 6, 50125 Firenze, Italy}
\author{M.~De~Rosa}
\affiliation{CNR-INO, Istituto Nazionale di Ottica,
Via Campi Flegrei 34, 80078 Pozzuoli (NA), Italy}

\begin{abstract}
We stabilize the idler frequency of a singly-resonant optical parametric oscillator directly to the resonance of a mid-infrared Fabry-Pérot reference cavity. This is accomplished by the Pound-Drever-Hall locking scheme, controlling either the pump laser or the resonant signal frequency. 
A residual relative frequency noise power spectral density below 10$^3$~Hz$^2$/Hz is reached, with a Gaussian linewidth of 920~Hz over 100~ms, which demonstrates the potential for reaching spectral purity down to the Hz level by locking the optical parametric oscillator against a mid-infrared cavity with state-of-the-art superior performance.

\end{abstract}


\maketitle


Stable, narrow-linewidth laser sources are essential for a wide range of demanding applications, including high-resolution spectroscopy, absolute frequency metrology, as well as production and manipulation of cold atoms and molecules~\cite{,Shelkovnikov:2008ht,Darquie:2010ho,Santamaria:2014bb,Salumbides:2015fh}. 
In this framework, a number of ambitious spectroscopic searches for new fundamental physics is underway; a favorable playground for these ultra-accurate measurements is the mid-infrared spectral region where most of the molecular ro-vibrational transitions exhibit natural linewidths in the 1-1000 Hz range.
However, while visible and near infrared lasers have already been narrowed down to sub-Hertz linewidths, mid-infrared sources have suffered some delay due to the lack of proper materials and devices.
Among the few exceptions, it is worth mentioning first the historical role played in the accurate measurement of the light speed by the CH$_4$-stabilized He-Ne laser at 3.39~$\mu$m; for this type of laser, linewidths below 10 Hz were obtained with sophisticated setups~\cite{Bagayev:1977bs}.
Extensive work has been made, in several decades, on CO$_2$ lasers as well, around 10~$\mu$m, leading to Hz-level spectral purity and sub-Hertz long-term stability~\cite{Bernard:1997td}.
More recently, quantum cascade lasers have come to the fore as versatile tools for precision spectroscopy, covering a vast part of the mid-infrared range up to the THz region and emitting up to a few hundred mW. They have already been narrowed to sub-kHz ~\cite{Taubman:02a,Cappelli:12a,Galli:13b} and sub-Hz linewidths~\cite{Argence:2015hc}. 
Following a twenty years long development, since the introduction of quasi-phase-matched crystals, optical parametric oscillators (OPOs) have proven to be reliable and versatile sources of continuous wave coherent radiation. They combine single-mode and high-power emission with tunability over a large spectral range, demonstrating sub-Doppler resolution in the mid-infrared~\cite{Kovalchuk:2001wo,Vainio:11c,Ricciardi:2012br}.
In the last decade, OPOs have been frequency stabilized against atomic or molecular absorptions and optical frequency combs, with the main goal of a long term stability and absolute frequency referencing~\cite{Zaske:10a,Vainio:11c,Kovalchuk:2005ju,Ricciardi:2012br,Peltola:2014eh}, or optical resonators, more effective on short time scales~\cite{Andrieux:2011uj,Mhibik:2011go,Ly:2015gn}. 
To date, narrowest observed linewidths are in the 10-100~kHz range \cite{Kovalchuk:2001wo,Ricciardi:12c,Ly:2015gn}.

In this Letter, we report frequency stabilization of the nonresonant idler wave of a mid-infrared OPO with respect to an external cavity, down to sub-kHz linewidth.
In our scheme, the frequency of the idler mode  is directly compared with a stable reference cavity according to the Pound--Drever--Hall (PDH) locking technique~\cite{Drever:1983gx}. 
An electronic servo converts the PDH signal into a control signal, which acts either on the pump frequency (pump-fed locking scheme, PF) or on the OPO cavity PZT (OPO-fed locking scheme, OF) in order to minimize the relative frequency noise.
  
\begin{figure}[t]
\begin{center}
\includegraphics*[bbllx=0bp,bblly=0bp,bburx=440bp,bbury=250bp,width=\columnwidth]{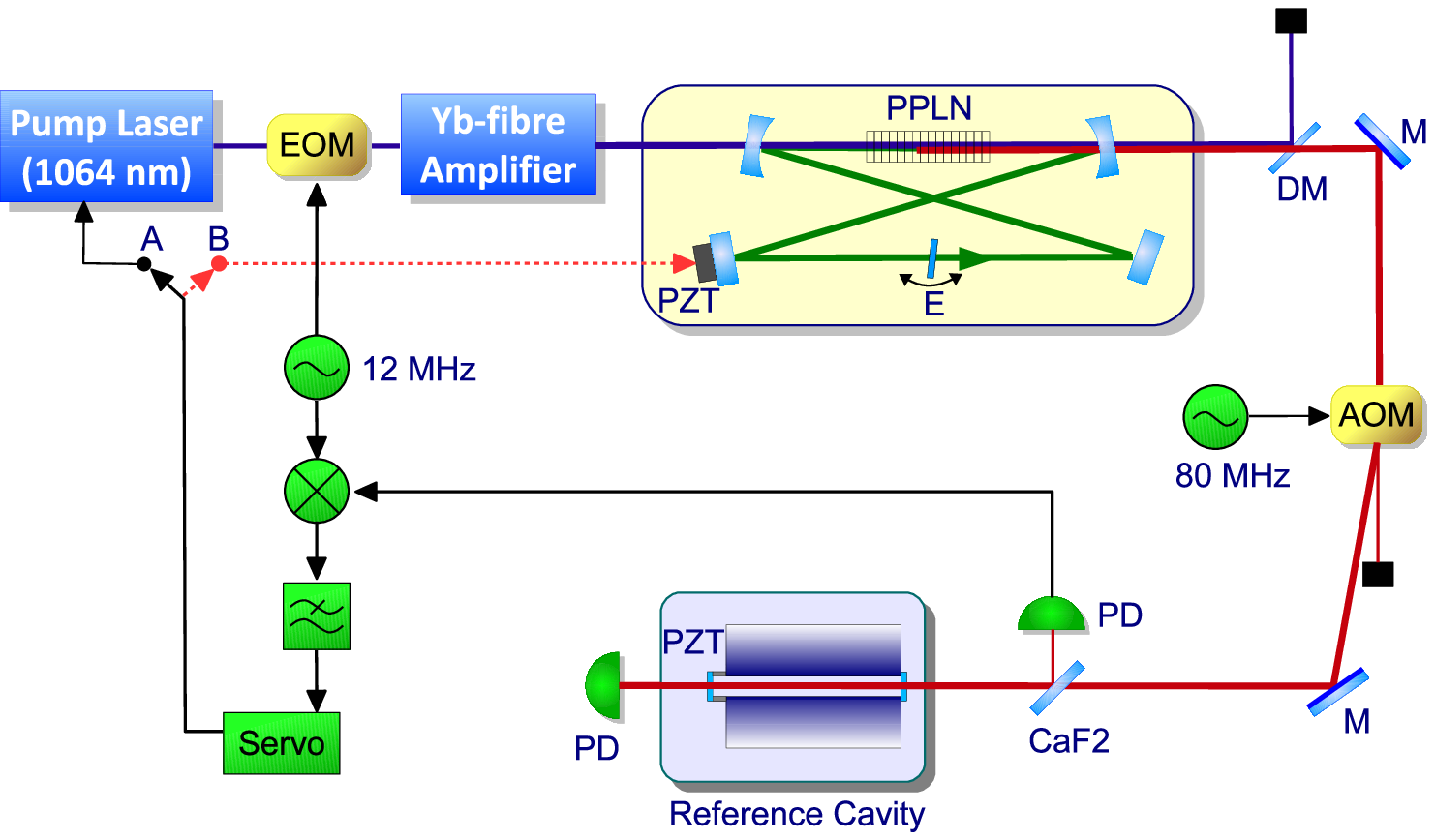}
\caption{Experimental scheme for frequency stabilization of the OPO idler mode. PPLN: periodically poled MgO:LiNbO$_3$ crystal; E: YAG etalon; PZT: piezoelectric actuator; EOM: electro-optic modulator; DM: dichroic mirror; M: mirror; AOM: acousto-optic modulator; PD: photodiode.}
\label{fig:setup}
\end{center}
\end{figure}

The experimental set-up is schematized in Fig.~\ref{fig:setup}.
The optical parametric oscillator is based on a periodically poled MgO:LiNbO$_3$ crystal placed in a bow-tie cavity resonant for the signal. A detailed description of the cavity is given in \cite{Ricciardi:2012br}.
The pump source is a Nd:YAG laser (Innolight, Mephisto), emitting 500~mW at 1064~nm, with a nominal linewidth of 1~kHz (over 100~ms). Frequency tuning is achieved, on a slow time scale ($\sim1$~s), by changing the laser crystal temperature and, on a fast time scale, by applying a high voltage signal on a PZT element acting on the crystal, with a response bandwidth of 100~kHz. About 10~mW of laser power seed an Yb-doped fiber amplifier and are amplified up to 10~W to pump the OPO. 
The nonlinear crystal has seven different poling periods, allowing continuous tuning of the idler frequency between 2.7 and 4.2 $\mu$m, with up to a Watt of emitted power. 
The overall performance has been improved with respect to the previous configuration. Indeed, the response bandwidth of the piezo-mounted mirror has been extended to about 50~kHz; for this purpose, we used a smaller and lighter mirror substrate in conjunction with a higher resonant frequency piezo (Piezomechanik, model PCh 150/5x5/2), mounted on a massive and lead-damped counterweight~\cite{Briles:2010tt}. In addition, we thermally stabilized the YAG etalon used for mode selection, enabling more robust long-term operation.
We also inserted an electro-optic modulator (EOM) in the pump beam optical path in order to add two sidebands at $f_{\rm PDH}=12$~MHz with respect to the laser carrier frequency; the pump frequency modulation is thus directly transferred to the OPO idler mode with the same modulation depth, avoiding the use of external modulation on the mid-infrared beam~\cite{Lindsay:2006tn}.

The reference cavity consists of two curved mirrors (1 m of radius of curvature) with a high-reflective coating (R > 99.7 \% ) in the range 3-3.5 $\mu$m, resulting in a measured finesse of $\sim1000$. The HR mirrors are mounted inside a stainless steel cell, kept under vacuum. In order to avoid back reflection, the idler beam is sent into an acousto-optic modulator (AOM) driven at $f_{\rm AOM}=80$~MHz, with about 50\% diffraction efficiency on the first order beam, that is then directed to the reference cavity. The incoming beam power coupled to the cavity $\text{TEM}_{00}$ mode is $>95$\%, indicating a very good spatial profile of the mid-infrared  radiation.
We tested our stabilization loop for different idler powers impinging on the reference cavity up to $\sim200$~mW. 
The radiation reflected from the cavity is partially deviated from a Ca$\rm F_2$ window towards a fast AC-coupled HgCdTe detector (Vigo System, model PV-4TE-5). The detected power is properly filtered and maintained to about 0.2~mW, irrespective of the idler power, in order to keep the signals well inside the detector dynamic range and fix the servo loop gain and parameters. The signal from the detector is then mixed with the local oscillator driving the EOM to retrieve the PDH signal, which is fed to the servo electronics.
The servo output provides two distinct correction signals: a ``slow'' control signal, effective for frequencies below 1 Hz, and a ``fast'' control signal, active for higher frequencies. In the PF scheme the slow signal acts on the crystal temperature of the Nd:YAG laser, while the fast signal drives the laser piezo. Alternatively, in the OF scheme the correction signals drive the OPO cavity PZT in a floating ground configuration: the slow signal is sent to a high-voltage amplifier and then to the PZT positive lead, while the fast signal is sent to the negative lead.
The correction and error signals are acquired and processed with a digital oscilloscope equipped with a built-in FFT routine to calculate the corresponding power spectral density (PSD).

\begin{figure}[t]
\centering
\includegraphics*[bbllx=5bp,bblly=0bp,bburx=300bp,bbury=230bp,width=\columnwidth]{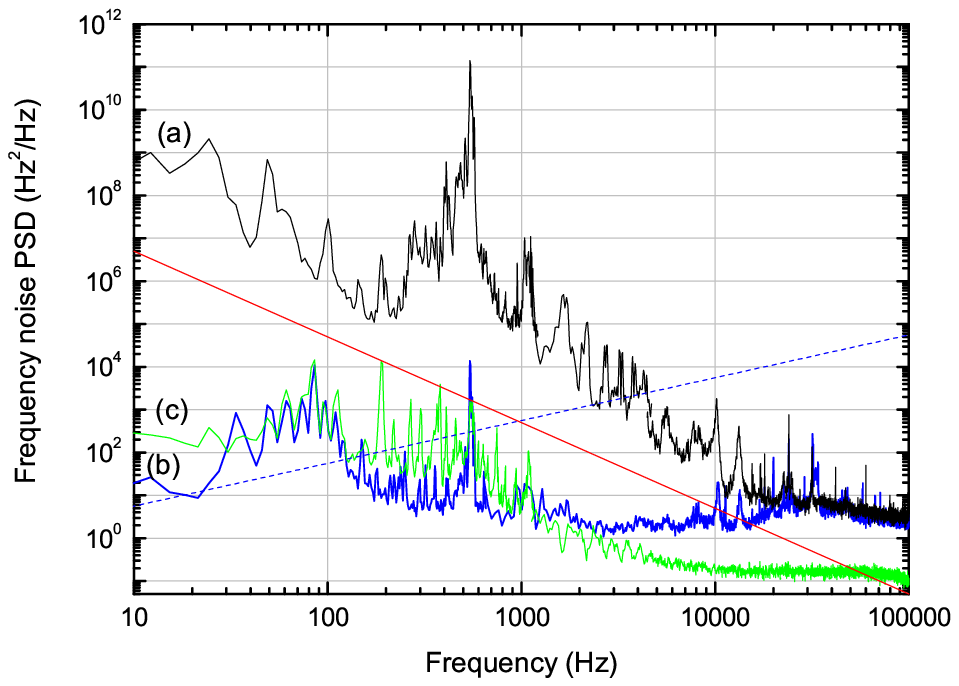}
\caption{Characterization of frequency noises. (a) Frequency noise PSD of the free running OPO. (b) In-loop frequency noise PSD for the idler mode in the PF stabilization scheme. (c) Detection limit. (continuous line) Pump laser $1/f^2$ frequency noise. (dotted line) $\beta$-separation.}
\label{fig:lockNdYAG}
\end{figure}

Figure~\ref{fig:lockNdYAG} reports the spectral characterization of frequency noise, in the PF stabilization scheme. Trace~(a) is the idler free-running frequency noise, obtained from the fast correction signal, while trace~(b) is the in-loop PSD obtained by the error signal, giving the residual relative frequency noise under active lock. The continuous $1/f^2$ line represents the pump laser frequency noise~\cite{Ricciardi:2013dn}. The peak around 500~Hz in trace~(a) has been identified as one of the mechanical resonances of the OPO cavity. In order to further reduce the mechanical noise, an OPO cavity with a more compact and robust design is under construction.
By comparing the in-loop and correction signal we estimate a total unit gain bandwidth of about 30~kHz, with a low frequency (10~Hz) noise reduction of almost eight orders of magnitude. Trace~(c) shows the detection limit, resulting by the FFT of the open-loop PDH signal, with the idler frequency far from a cavity resonance. At high frequencies the detection limit is set by the detector dark noise, which is equivalent to a frequency noise level $<1~$Hz$^2$/Hz, while the excess noise at frequencies below 1~kHz is partly due to a residual idler amplitude noise and to interference fringes. 
The  $\beta$-separation (dotted line), given by $\beta(f)=8\text{ln}(2)f/\pi^2$, visually separates the spectrum in two regions which have a different impact on the emission line shape~\cite{DiDomenico:2010tx}: where the noise level is higher than the $\beta$-separation, the spectrum contributes to the central part of the line shape and thus to the linewidth, whereas the spectrum portion lying under the $\beta$-separation mainly contributes  to the wings of the line shape but does not affect the linewidth. 

\begin{figure}[t]
\centering
\includegraphics*[bbllx=5bp,bblly=0bp,bburx=300bp,bbury=230bp,width=\columnwidth]{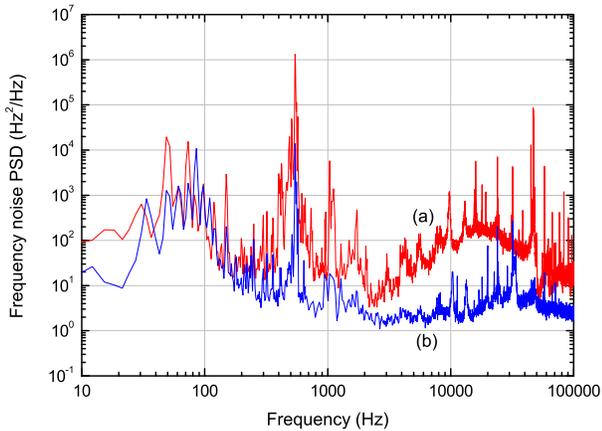}
\caption{OPO residual frequency noise: comparison between OF and PF scheme. (a) In-loop frequency noise PSD relative to the OF scheme. (b) In-loop frequency noise PSD relative to the PF scheme (trace (b) in Fig.~\ref{fig:lockNdYAG}).}
\label{fig:lockOPO}
\end{figure}

Figure~\ref{fig:lockOPO} compares the residual frequency noise obtained (a) in the OF scheme and (b) in the PF scheme. In the former case, the locking bandwidth is limited to about 10~kHz, because of the reduced frequency response of the OPO piezo actuator with respect to the laser one. The reduced bandwidth results in a less effective noise reduction in the OF scheme with respect to the PF scheme.

We also characterized the spectral emission in terms of linewidth. From the PSDs of the correction and error signals we calculated the idler line shape using the Elliott's formula~\cite{Elliott:1982di} and retrieved the corresponding linewidth in both the free-running and locked configuration. 
According to the $\beta$-separation criterion, for both the locking schemes, the spectral frequencies that significantly contribute to the linewidth lie in the 10-100 Hz range, with several isolated peaks around 500 Hz. This analysis suggests an ``observation time'' of 100~ms, over which the line shape calculation is meaningful.
 
\begin{figure}[t]
\centering
\includegraphics*[bbllx=5bp,bblly=0bp,bburx=300bp,bbury=240bp,width=\columnwidth]{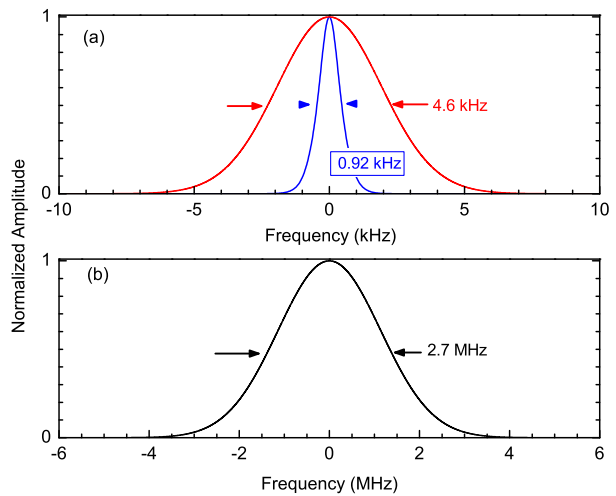}
\caption{ (a) Calculated line shapes of the stabilized OPO on 100 ms time scale: PF scheme (0.92~kHz linewidth), OF scheme (4.6~kHz linewidth). (b) Calculated line shape of the free-running OPO on 100 ms time scale (2.7~MHz linewidth). Notice the different frequency scales of the horizontal axes. 
}
\label{fig:linewidth}
\end{figure}

Figure~\ref{fig:linewidth}(a) shows the line shapes relative to the PF and OF schemes, calculated for a time scale of 100~ms, yielding linewidths of 0.92~kHz and 4.6~kHz, respectively. The line shape of the free running OPO, calculated on time scale of 100~ms, provides an idler linewidth of 2.7~MHz, Fig.~\ref{fig:linewidth}(b). All the above line shapes are well fitted by Gaussian profiles.
Finally, we notice that most of the free running OPO frequency noise relevant for the linewidth lies below a few~kHz; in fact, the line shape calculation on a shorter time scale (1~ms) results in a linewidth of 52~kHz.

Summarizing, we demonstrate relative frequency stabilization of the mid-infrared idler mode of a singly resonant OPO with respect to a medium-finesse reference cavity, obtaining a sub-kHz relative linewidth on a 100-ms time scale. 
The present limitations on the attainable frequency noise level are mostly due to intrinsic cavity noise and environmental disturbances, as the detection limit set by the detector dark noise is well below $1$~Hz$^2$/Hz and  the quantum limit set by the shot noise~\cite{Black:2001tt}, for an impinging power of 0.2~mW on the detector, is about $3\cdot10^{-6}~$Hz$^2$/Hz.
We remark that a higher PDH signal slope, as can be obtained with a higher cavity finesse, would further lower the limit due to dark and shot noise. Nevertheless, the achieved result, more than an order of magnitude better than what has been done so far \cite{Ly:2015gn}, is an important step towards the development of a widely tunable, powerful mid-infrared laser source which may simultaneously exhibit a sub-Hz linewidth through the use of a highly stable reference cavity.
In this respect, a long standing effort has been addressed to improve the frequency stability of Fabry--P\'erot cavities~\cite{Notcutt:2005hi}, reaching record performances in terms of fractional stability, $10^{-16}$, and linewidth, 40~mHz~\cite{Kessler:2012gm}. 
However, dielectric mirrors used in Fabry--P\'erot cavities have a bandwidth typically limited to a few hundreds nm, thus requiring different sets of mirrors to cover the whole tunability range of our OPO.
To overcome this limit, an interesting solution is the use of monolithic whispering-gallery-mode resonators~\cite{Alnis:2011bi}, where the light mode is confined by total internal reflection and the useful spectral range extends therefore to the whole transparency window of the material, thus covering the entire OPO emission range. Since thermorefractive fluctuations set a limit on the achievable frequency noise level, microresonator material and operating temperature require a careful choice. Useful materials in the emission range of our OPO include dielectric crystals, such as MgF$_2$, Al$_2$O$_3$, or semiconductors, like Si, Ge, AlN, GaAs etc.~\cite{Schliesser:2012dn,Wang:2013db}.
Finally, to compensate for long term drift, the reported locking scheme can be combined, e.g., with a frequency comb referencing, ensuring frequency stability on long time scales and additionally providing an absolute frequency calibration.  

\bibliography{OPOlocking}

\end{document}